# Progress in Developing Hybrid RPCs: GEM-like Detectors with Resistive Electrodes

P. Fonte[1], P. Martinengo[2], E. Nappi[3], R. Oliveira[2], V. Peskov[2,4]*

[1]ISEC and LIP, Coimbra, Portugal
[2]CERN, Geneva, Switzerland
[3]INFN Bari, Bari, Italy,
[4]Ecole Superior des Mines in St Etienne, France

**Abstract**

We have recently developed an innovative detector of photons and charged particles: a GEM-like gaseous amplification structure with resistive electrodes instead of commonly used metallic ones. This novel detector combines the best property of GEMs- the capability to operate in a cascaded mode and in poorly quenched gases - and of RPC: the protection against sparks.
In this paper will shortly review our latest achievements in this direction, however the main focus will be given on a new advanced design that allows to build large area detectors  manufactured by a screen printing technology. The proposed detector, depending on the applications, can operate either in a GEM mode (electron multiplications through holes only) or as a hybrid RPC with simultaneous amplifications in the drift region and in the holes. The possible applications of this new detector will be discussed.

## I. Introduction

In our recent papers [1-5], we have described GEM-like detectors with electrodes made of resistive materials instead of commonly used metallic ones. We named these new type of detectors: Resistive Electrode Thick GEMs or RETGEMs. At low counting rates they operate as do usual GEMs, however in the case of discharges which may appear at high gains due to the high resistivity of the electrodes, the energy released by the sparks is rather small making these detectors intrinsically spark protected. Therefore, from this point of view they behave like RPCs.
The electrodes of the first RETGEM prototypes [1] were manufactured from a resistive graphite-based paint (~100μm thick) usually used for manufacturing the outer electrodes of RPCs (see for example [6]). The surfaces between the holes were covered with this layer by a tiny brush (several coatings were applied) making the procedure very demanding.
Later on, a more advanced technology in the RETGEM manufacturing was implemented: the resistive electrodes were made of resistive Kapton (type 100XC10E5) glued on top of the G-10 surface [2]. These detectors were easier to make and all of them had a high quality of holes and edges around the holes thus allowing to reach exceptionally high gas gains of ~$10^5$ for 6 keV photons.

---

Unfortunately, quite recently, some export restrictions have been introduced for the use of the DuPont Kapton 100XC10E5 outside the U.S.A., therefore we tried to find alternatives technologies.

In [5] we have described the first results obtained with RETGEMs manufactured by a screen printing technology. The advantage of this technique is that it is widely accessible in many research labs and companies. The screen printing technology also allows one to also manufacture RETGEMs of various resistivities optimized for particular experimental conditions or practical applications.

One should note that the above described RETGEMs were just simplified prototypes used for studies of the features of their operation and they were certainly not optimized for any practical application. For example, the HV was applied to the resistive electrodes via Cu frames manufactured in the periphery of the detector. With such geometry the avalanche current should finally flow to this frame through the entire resistive layer which creates undesirable voltage drops along its surface and in some cases even causes surface streamers between the holes and the Cu frame [2]. To avoid the surface streamers, in improved prototypes the Cu frame was placed at some distance (~1 cm) from the edge of the active area of the detector (the area containing holes). As a consequence of this modification, the ratio of the active area A to the full detector area F was typically A/F ~30%. Thus the designs with charge collecting frames were not suitable for large area detectors.

The aim of the present activity is to develop industrial processes for building large-area RETGEMs or detectors consisting of a mosaic of small detectors characterized by high ratios A/F.

## II. Detector Design and Experimental Set-up

We have developed doubled layered RETGEMs with an inner layer consisting of a metallic mesh or a grid and an outer layer made of resistive material manufactured by a screen printing technology.

Fig. 1 shows the main steps in such a detector's manufacturing process. First of all, on both surfaces of the G-10 plate a Cu mesh (Fig.1a) or a Cu grid (Fig.1b) was manufactured by a photolitographic technology. In both designs, the thickness of the strips was 30 μm and their height was 13 μm. On the top of these electrodes a 15 μm thick resistive layer was formed by a screen printing technology. These double layered plates were then cured in the oven at temperatures of 200°C and after that, holes were drilled in between the strips. Their diameter, depending on the particular design, was in the range of 0.2-0.5 mm and the pitch from 0.4 to 0.7 mm, correspondingly. RETGEMs with active areas of 5x5 cm$^2$, 10x10 cm$^2$ were manufactured. The detector with the inner mesh layer was called the "M- REGEM", and the detector with the grid structure was called the "G- RETGEM". We also manufactured one 5x5 cm$^2$ G-RETGEM with resistive strips on the top of metallic grid, so that the resistive layer was segmented. In this design the resistive grid covered a group of metallic strips, two in each group. We called it the S-RETGEM.

In contrast to the previous RETGEM prototypes, the new designs allow one to collect avalanche charges on the strips close to the hole where the avalanche develops and thus to minimize the current flow along the surface. Moreover, the grid design offers the possibility to obtain information about the avalanche positions directly from the G-RETGEM's electrodes via the signals measured from its strips.

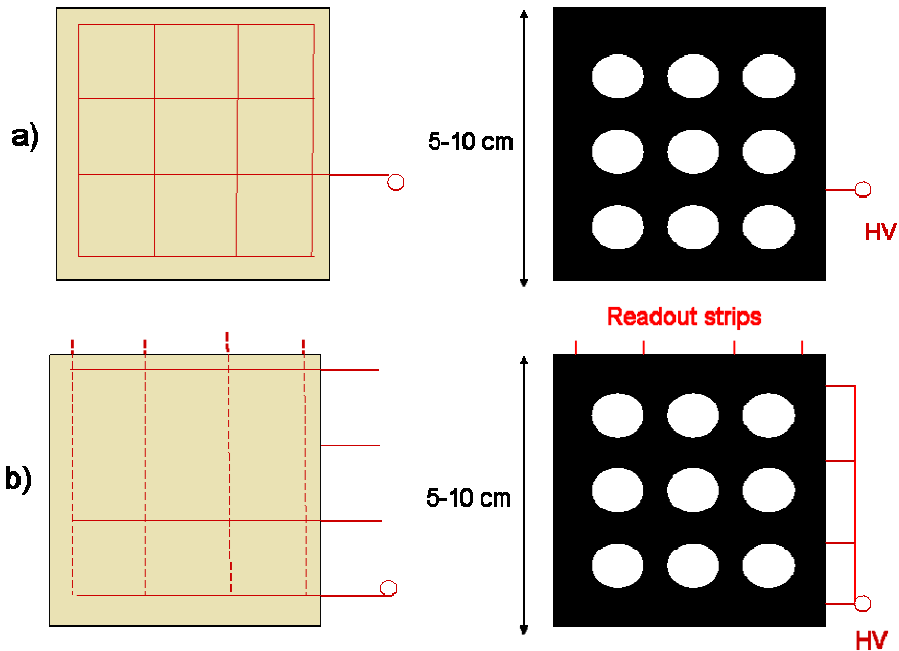

Fig. 1. Main steps in detector's manufacturing: a) M-RETGEM, b) G-RETGEM; the dash lines show the grid manufactured on the other side of the G-10 plate.

The quality of the manufactured detectors was preliminarily evaluated by measuring the breakdown voltage $V_{air}$ in air. If $V_{air} > 3$ kV, the detector was accepted for further tests. If $V_{air} < 3$ kV, the detector was additionally treated by a sand jet technique allowing one to remove imperfections inside the holes and around their edges. Usually, after such treatment, the value of $V_{air}$ increases up to 3 kV or even more so that finally such detectors were considered as having good quality.

Further tests were performed with the set-up shown in Fig 2. It consists of a gas chamber housing a single RETGEM or two RETGEMs operating in a cascaded mode. A gas system allowed to flush various gases (Ne, Ar or Ar+5%$CO_2$).

The ionization inside the gas was produced by alpha particles emitted from $^{241}$Am or by 6 keV photons from $^{55}$Fe.

In the case of the M- RETGEM, the signals delivered by the electrodes were detected either by a charge sensitive or by a current amplifier, depending on the specific measurements. In the case of the G-RETGEM several current amplifiers were connected to the central strips whereas the adjacent strips were either grounded or connected to 100 kΩ resistors (to study if there is any signal crosstalk).

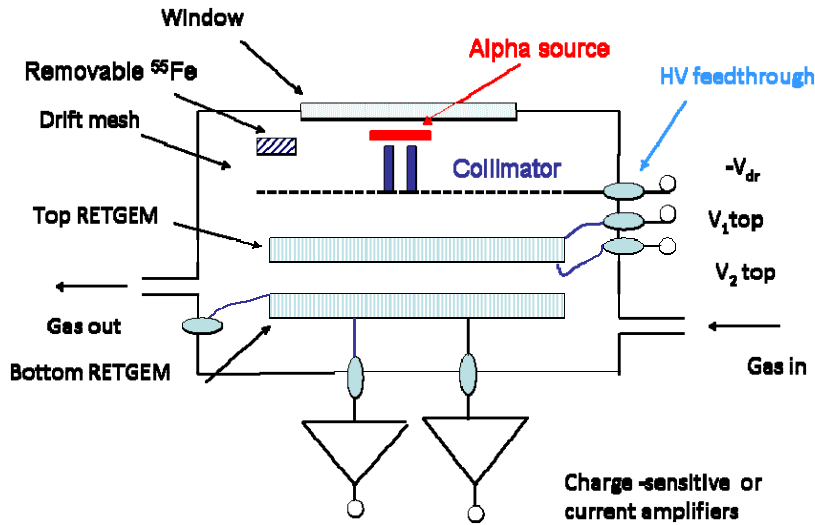

Fig. 2. Schematic drawing of the experimental set-up.

In the S-RETGEM there was no avalanche charge leak between the metallic strips (belonging to different hole's row) via the segmented resistive layer so this detector was readout by charge sensitive amplifiers.
For comparative studies we also used an "old RETGEM" prototypes manufactured by a screen printing technology and containing a charge collecting Cu frame [5].
In additional measurements of spark energy released in various detectors and various gases, we also used thick GEMs with metallic electrodes (TGEMs) [4] and TGEMs with 15 μm thick resistive coatings manufactured by a screen printed technology on the top of their Cu electrodes. These detectors had the same geometry as the RETGEMs described above.

## III. Results

Fig. 3 shows the gain vs. voltage curves measured for both M- and G-RETGEMs. The maximum achievable gain of these detectors was almost the same as we usually achieve with the old RETGEMs [5].
We also tested the operation of M- and G-RETGEMs with gas pre-amplification in the drift region. This mode of operation, first suggested in [7], may offer better position [7,8] and time resolutions and could be very attractive for some applications, for example in the case of triggered RETGEMs [9]. Typical results are presented in Fig. 4. One can see that 2-3 times higher gains were achieved in this operation mode.

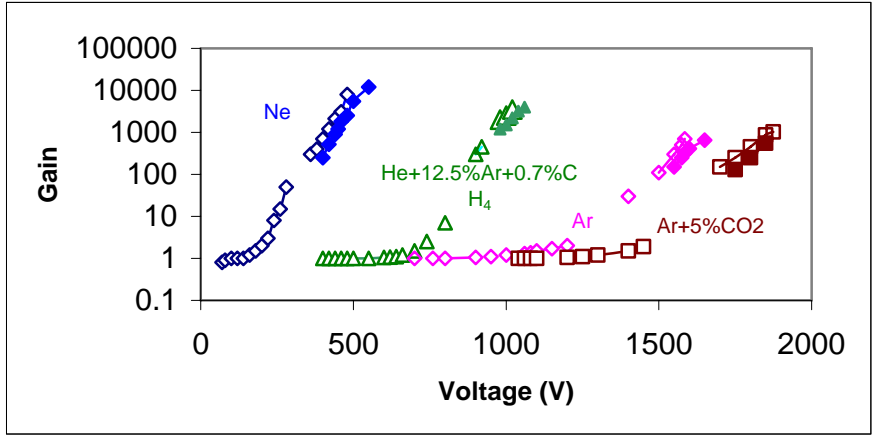

Fig. 3. Gain vs. voltage for M-RETGEM (open symbols) and G-RETGEM (filled symbols) measured with alpha particles (curves with gains less than 100) and with 6 keV photons (curve with gains higher than 100) in various gases.

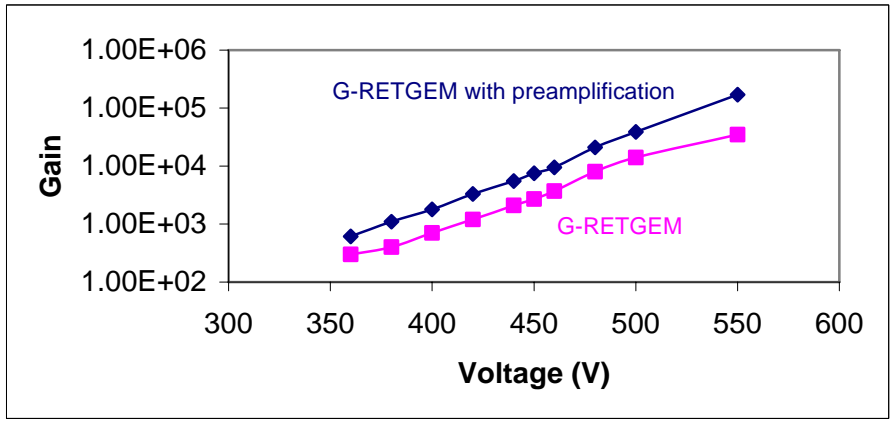

Fig. 4. Gain curves of G-RETGEM and G-RETGEM operating with gas pre-amplification in the drift region measured in Ne.

Another set of tests were done to demonstrate that, in the case of G-RETGEMs, the information about the avalanche position could be obtained from the measurements of the strips signals. As an example, Fig. 5 shows the signal amplitudes vs. the strip number measured by current amplifiers for the following two experimental arrangements: 1) the area between the strip # 10 and #11 of the G-RETGEM was irradiated by a beam (diameter of 0.5 mm) of collimated alpha particles, 2) the beam of alpha particles was collimated in the middle of the strip #10. One can see that these curves show a sharp maximum at the strip's numbers corresponding to the irradiated area thus indicating that a position resolution of about the pitch of the detector can be obtained by this method.
Similar measurements were performed with S-RETGEMs using charge sensitive amplifiers (see Fig. 5). The advantage of this design is that position measurements can be performed at much smaller gains than in the case of the G-RETGEM readout by the current amplifiers.

In the near future, we will perform more accurate measurements with a better collimated beam of soft x-rays.

Fig. 6 shows the rate characteristics of the M- and G-RETGEMs and, for comparison, the rate characteristics of the old RETGEM (with a Cu frame). One can see that due to better charge collection on the metallic electrodes placed under the resistive layer, the improved detectors (with the mesh or grid) show much better rate characteristics.

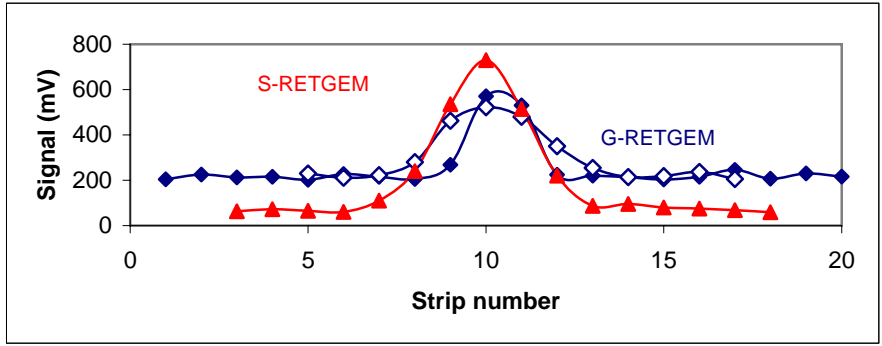

Fig. 5. Signal vs. strip number measured with G- and S-RETGEMs with the help of alpha particles collimated in between the strips #10 and #11(filled symbols) and collimated in the middle of the strip #10 (open symbols).

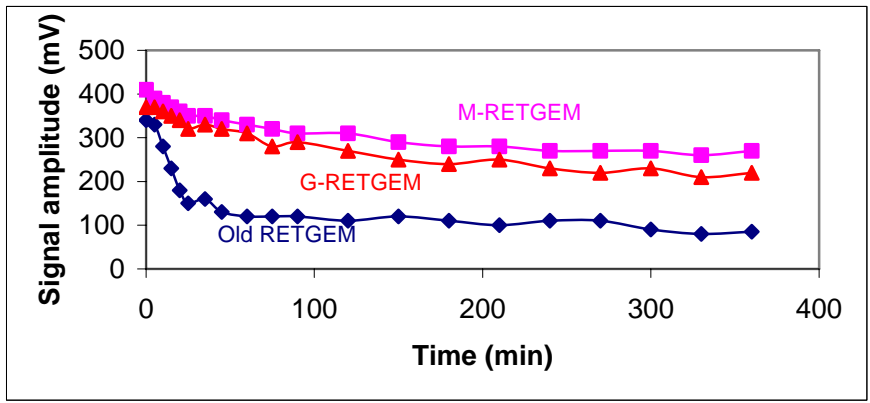

Fig. 6. Signal amplitude vs. time for old RETGEM, M-RETGEM and G-RETGEM when the detectors were irradiated by 6keV photons (~450 Hz/cm$^2$). Measurements were performed in Ne at gas gains of $10^4$.

Further studies were performed to verify if the new designs provided efficient spark suppression. For this purpose, the sparks current was measured using M-RETGEMs, G-RETGEMs, TGEMs and TGEMs coated by the screen printed technology with restive layers. The results obtained were similar to the ones described in [3]: the discharge current in the TGEM with a resistive coating was typically 4-7 times less than in the case of the TGEM, whereas the discharge current in the mesh and grid RETGEMs was ~500 times less than for the TGEMs of the same geometry. Thus, both new designs provided as a good spark suppression as the old RETGEMs did.

## IV. Conclusions

Both new proposed designs, based on inner metallic mesh and grid, allowed to manufacture large area RETGEMs. Detectors with an active area of 10x20 cm$^2$ were already developed recently. In principle, the screen printing technology is suitable to build even much larger detectors, up to 50x100 cm$^2$.

In order to obtain a position information from the M-RETGEM, one has to extract the avalanche charge from the holes to the readout plate as it is the case for all hole-type detectors, including GEMs (see for example [10]).

In this work for the first time we have experimentally demonstrated that, in the case of the G-RETGEM, one can obtain the information about the avalanche position from the measurements of the strip signals. This approach has several advantages, very relevant to some applications, for example such a detector in a final assembly has less elements (no readout plate) and it can be operated at lower gas gains since there is no charge extraction from the holes (usually only 25-50 % of the avalanche charge is collected on the readout plate) and it can operate in electro-negative gases. Moreover, with the new designs it is possible to exploit the pre-amplification process in the drift region which may offer better position and time resolutions.

We believe that the new RETGEMs will find a lot of applications. Our group, for example, suggests to use M- and G- RETGEMs in the upgrade of the ALICE RICH detectors, called VHMPIDs [9, 11].